# On the Optimum Scenarios for Single Row Equidistant Facility Layout Problem

SHROUQ GAMAL[*]
SHROUQGAMAL@f-eng.tanta.edu.eg
Ahmed A. Hawam[*]
Ahmed M. El-Kassas[*]
[*]Production Engineering and Mechanical Design Department, Faculty of Engineering,
Tanta University, Egypt

**Abstract**

In this paper, we search Single Row Equidistant Facility Layout Problem SREFLP which is with an NP-Hard nature to mimic material handling costs along with equally spaced straight-line facilities layout. Based on the literature, it is obvious that efforts of researchers for solving SREFLP turn from exact methods into release the running time tracing the principle of the approximate methods in time race, regardless searching their time complexity release in conjunction with provable quality of solutions.  This study focuses on Lower bounding LB techniques as an independent potential solution tool for SREFLP. In particular, best-known SREFLP LBs are reported from literature and significantly LBs optimum scenarios are highlighted. Initially, one of the SREFLP bidirectional LB gaps is enhanced. From the integration between the enhanced LB and the best-known Gilmore-Lawler GL bounding, a new SREFLP optimum scenario is provided. Further improvements to GLB lead to guarantee an exact Shipping/Receiving Facility assignment and propose a conjecture of at most 4/3 approximation scheme for SREFLP.

## 1. Introduction

In almost any Facility Layout Problem one can think of, productivity is best served by an efficient flow of the elements that move through the facilities. It might be said that the overall success of a Layout, or at least its profitability, is a direct reflection of the effort that goes into the flow planning. Accordingly, it is reflected on material handling cost minimization. The solution key in Facility Layout and Material Handling Problems is to form optimum effective facilities sequence/s. If all the facilities are assigned on equally spaced same side of the material-handling track, the layout is referred to as a Single Row Equidistant Facility Layout SREFLP. A literature review on SREFLP enormous applications can be found in detail by (Palubeckis, 2012).

SREFLP is a well-known operations research problem with a permutation-based manner. A key issue of SREFLP is its NP hard nature. That means there is no polynomial time algorithm for solving it exactly unless P = NP. To out of this dilemma, literature sequentially tackled SREFLP Problem exactly then heuristically. We consider only the four recent contributions of SREFLP approaches to be reviewed: lower bounds-based exact algorithms (Hungerländer, 2014; Palubeckis, 2012) and the best-known solutions using approximate-based methods (Atta & Sinha Mahapatra, 2019; Palubeckis, 2015) as follows: In (Palubeckis, 2012), the specialized (LP)-based approach, using the best-known B&B lower bound LB* of SREFLP, was able to solve instances size up to 35 facilities. Recently, (Hungerländer, 2014) reported the superior results of applying (SDP)- based approach of the general Single Row Facility Layout Problem SRFLP, which was proposed by (Hungerländer & Rendl, 2011) to solve SREFLP instances up to 42 facilities. Reference (Hungerländer, 2014) computationally improved LB* tightness (Palubeckis, 2012) to emerge it in (SDP)-based approach for heuristically solving larger instances. However, it seems that improved LB* is not the absolute condition for improving the quality of the proposed heuristic. As observed in results of SREFLP instances Y with size 45-60 departments (Hungerländer, 2014), best layouts are obtained by LB* of (Palubeckis, 2012) which is weaker than the improved LB* of (Hungerländer, 2014) For improvement the quality of solution, (Hungerländer, 2014) proposed a note to treat this gap using his improved LB* for the specialized (LP)-based approach of (Palubeckis, 2012). Although no computational results reported this procedure till now, it is expected either no running time release with approximate solutions or running time release with again approximate solutions.

In this context, (Palubeckis, 2015) applied proposed simulated annealing SA algorithms for the same instances Y with size 45- 60 departments. It conjectured that SA algorithms produced optimal results with less computational effort; it is unproven. Reference (Atta & Sinha Mahapatra, 2019) is considered the recent study on SREFLP results. Similarly, as derived in (Atta & Sinha Mahapatra, 2019; Hungerländer, 2014) tailored a SRFLP heuristic to apply on SREFLP which outperformed on computational time while agreed with all optimum results of (Hungerländer, 2014; Hungerländer & Rendl, 2011; Palubeckis, 2012) that is up to 35 facilities. The running time of (Atta & Sinha Mahapatra, 2019) is comparable to (Palubeckis, 2015) for agreed results of instances ($110 \leq n \leq 300$). Regarding the literature, it is obvious that efforts of researchers in exact methods turn into release the running time tracing the principle of the approximate methods in time race, regardless searching their time complexity release in conjunction with provable quality of solutions.

To show that, Enhanced LB is initially proposed. Based on the literature, the only two mentioned optimum LB scenarios are cited. The implicit object of this study is to mine the SREFLP LB literature, reanalyze it, and highlight every possible scenario in support of solving SREFLP exactly and polynomially (conjecture P=NP). To support that, this paper introduced:

1. A new Optimum Scenario for SREFLP LB resulted from the possible integration between the best-known Gilmore Lawler bound and the Enhanced LB.

2. Exact Shipping/Receiving Facility Assignment generated from partially considering GLB as a solution-tool.
3. Conjecture on a 4/3 fully input polynomial time approximation scheme for SREFLP related to fully consider GLB as a solution-tool taking into account the exact assignment of shipping/ receiving facility.

## 2. Single Row Equidistant Facility Layout Problem

SREFLP can be considered as a Frequency From-To Chart FFTC Problem reduced into a Frequency From-Between Chart FFBC through the following structure from (Gamal et al., 2020) and (Palubeckis, 2015), respectively:

$$\text{Min FFTC} = \sum_{i=1}^{n-1} \sum_{j=i+1}^{n} (f_{ij} + f_{ji})(j - i), \quad (1)$$

$$\min_{\pi \in \mathfrak{n}} \text{FFBC} = \sum_{i=1}^{n-1} \sum_{j=i+1}^{n} f^{t}_{\pi(i)\pi(j)} (j - i), \quad (2)$$

where $\mathfrak{n}$ is a set of all $n!$ permutations. For $i \neq j$, $i = 1, 2, .., n-1$. and $j = 2, 3, .., n$. $f^{t}_{\pi(i)\pi(j)}$ represents the Forwards and Backtracks frequencies sum $(f_{ij} + f_{ji})$ of the material flow between facilities $i$ and $j$. Term $(j - i)$ states the distance between facilities $i$ and $j$. Equation (2) may be referred as $f^{t}_{ij_{(j-i)}}$ (see Fig.1).

| Between From | Facility 1 | Facility 2 | Facility 3 | Facility 4 | Facility 5 |
|---|---|---|---|---|---|
| Facility 1 | 0 | $f^{t}_{12_1}$ | $f^{t}_{13_2}$ | $f^{t}_{14_3}$ | $f^{t}_{15_4}$ |
| Facility 2 | 0 | 0 | $f^{t}_{23_1}$ | $f^{t}_{24_2}$ | $f^{t}_{25_3}$ |
| Facility 3 | 0 | 0 | 0 | $f^{t}_{34_1}$ | $f^{t}_{35_2}$ |
| Facility 4 | 0 | 0 | 0 | 0 | $f^{t}_{45_1}$ |
| Facility 5 | 0 | 0 | 0 | 0 | 0 |

Fig. 1 Frequency From-Between Chart Skeleton

SREFLP could be categorized as:

- Full SREFLP, whose all $(n(n-1)/2)$ total frequencies cells are non-zeros. It practically means that every facility has at least one bi/directional flow among all other $(n-1)$ facilities (see Fig.2a.).
- Non-full SREFLP, whose some of $(n(n-1)/2)$ total frequencies cells contain zeros. It practically means that every facility hasn't at least one bi/directional flow among all other $(n-1)$ facilities (see Fig. 2b.).

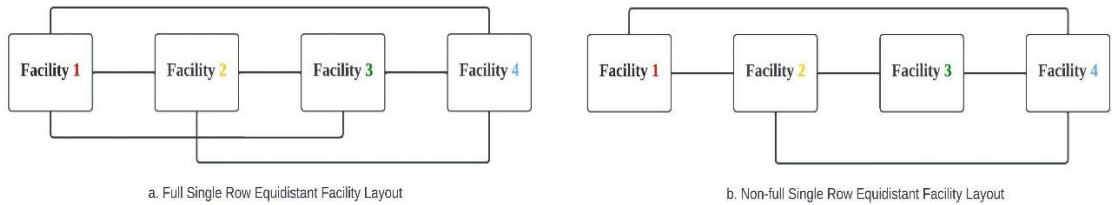

a. Full Single Row Equidistant Facility Layout                b. Non-full Single Row Equidistant Facility Layout

Fig. 2 Single Row Equidistant Facility Layout

## 3. SREFLP Lower Bounding Techniques

In most cases of SREFLP, LB is used as an aid tool for reaching exact solutions or as a basis for measuring the performance of heuristic approaches. That is through either computing infeasible bound or may optimum one; it is unprovable.

In the following sections, the four LB related to SREFLP are discussed and reanalyzed in support of the conjecture P=NP.

### 3.1 Enhancement of SREFLP Bi-directional Lower Bound

From literature, Initially, we enhance a previous version of Frequency From-To Chart FFTC computing Bi-directional LB for SREFLP. It was suggested by (Sarker et al., 1998) and continued to be used as a measuring tool of the heuristic performance in (Sarker, 2003). It claimed an effect of Backtracking Minimization over Moment Minimization, which has been numerically proven to be incorrect by (Gamal et al., 2020).

The Enhanced LB computation is based on the skeleton of Objective Function (1), as follow:

> In general, for $n \times n$ FFBC, setting the frequencies set $f_{ij}^t$ of $n(n-1)/2$ non-diagonal cells $\{f_N^t\}_{N=1}^{n(n-1)/2}$ in descending order array, i.e., such that $f_N^t \geq f_{N+1}^t$ for all $N$. Then, the distance array is provided by $1< k < n+1$. Finally, Enhanced LB Model can be formulated as:
>
> $$\text{Enhanced LB} = \text{Sum}\left[f_1^t \ f_2^t \ .. \ f_{\frac{n(n-1)}{2}}^t\right] \cdot \left[\underbrace{1 \ 1 \ .. \ 1}_{n-1 \text{ times}} \ .. \ \underbrace{k \ k \ .. \ k}_{n-k \text{ times}} \ .. \ n-2 \ n-2 \ n-1\right] \quad (3)$$

From the following Comparative Example and Experimental Example, Enhanced LB is expected to be an infeasible solution or Optimum LB, which is an optimum solution; it is unproven.

#### 3.1.1 Comparative Example

Illustration of Enhanced LB for bidirectional flow (Moment Minimization) problem versus Bi-directional LB computed in (Sarker et al., 1998).

Consider a From-To Chart FTC matrix for a 5-machine problem:

$$\text{FTC} = \begin{bmatrix} 0 & 2 & 2 & 3 & 0 \\ 1 & 0 & 5 & 2 & 3 \\ 6 & 1 & 0 & 4 & 2 \\ 1 & 2 & 6 & 0 & 3 \\ 1 & 5 & 3 & 3 & 0 \end{bmatrix}$$

Regarding (Sarker et al., 1998) : Forward Lower Bound is computed by firstly choosing the larger frequencies between every pair of $(f_{ij}, f_{ji})$ such that $\{(2,6,3,1), (5,2,5), (6,3), (3)\}$, then placing in an array with ascending order: [1 2 2 3 3 3 5 5 6 6]. And the distance array is placing in descending order: [4 3 3 2 2 2 1 1 1 1]. Thus, Forward Lower Bound equals to 56 by taking the product of both arrays.

Backtracking Lower Bound is computed by firstly choosing the smaller frequencies between every pair of $(f_{ij}, f_{ji})$ such that {(1,2,1,0), (1,2,3), (4,2), (3)}, then placing in an array with ascending order: [0 1 1 1 2 2 2 3 3 4]. And the distance array is placing in descending order: [4 3 3 2 2 2 1 1 1 1]. Thus, Backtracking Lower Bound equals to 28 by taking the product of both arrays.

Hence, Bi-directional LB = Forward Lower Bound + Backtracking Lower Bound = 84.

Regarding Enhanced LB:

$$\text{From-Between Chart FBC} = \begin{bmatrix} 0 & 3 & 8 & 4 & 1 \\ 0 & 0 & 6 & 4 & 8 \\ 0 & 0 & 0 & 10 & 5 \\ 0 & 0 & 0 & 0 & 6 \\ 0 & 0 & 0 & 0 & 0 \end{bmatrix}$$

$\{f_N^t\}_{N=1}^{10} = \{3, 8, 4, 1, 6, 4, 8, 10, 5, 6\}$

$f_{(N)descend}^t = [10\ 8\ 8\ 6\ 6\ 5\ 4\ 4\ 3\ 1]$

Distance array = $[1\ 1\ 1\ 1\ 2\ 2\ 2\ 3\ 3\ 4]$

Enhanced LB = Sum $[10\ 8\ 8\ 6\ 6\ 5\ 4\ 4\ 3\ 1] \cdot [1\ 1\ 1\ 1\ 2\ 2\ 2\ 3\ 3\ 4]$ = 87

Hence, Bi-directional LB equals 84 (Sarker et al., 1998) < 87 (Enhanced LB) < 90 (Optimal solution). Thus, Enhanced LB outperforms Bi-directional LB and produces infeasible solution.

### 3.1.2 Experimental Example

Consider a From-To Chart FTC matrix for a 5-machine problem:

$$\text{FTC} = \begin{bmatrix} 0 & 2 & 2 & 1 & 1 \\ 0 & 0 & 1 & 1 & 1 \\ 0 & 1 & 0 & 4 & 2 \\ 0 & 1 & 0 & 0 & 2 \\ 0 & 0 & 0 & 0 & 0 \end{bmatrix}$$

Regarding Enhanced LB:

$$\text{FBC} = \begin{bmatrix} 0 & 2 & 2 & 1 & 1 \\ 0 & 0 & 2 & 2 & 1 \\ 0 & 0 & 0 & 4 & 2 \\ 0 & 0 & 0 & 0 & 2 \\ 0 & 0 & 0 & 0 & 0 \end{bmatrix}$$

$\{f_N^t\}_{N=1}^{10} = \{2, 2, 1, 1, 2, 2, 1, 4, 2, 2\}$

$f_{(N)descend}^t = \{4, 2, 2, 2, 2, 2, 2, 1, 1, 1\}$

Distance array = $[1\ 1\ 1\ 1\ 2\ 2\ 2\ 3\ 3\ 4]$

Enhanced LB = Sum $[4\ 2\ 2\ 2\ 2\ 2\ 2\ 1\ 1\ 1] \cdot [1\ 1\ 1\ 1\ 2\ 2\ 2\ 3\ 3\ 4]$

= 32 (OPT-LB)

According to (Experimental Example), Enhanced LB may generate OPT-LB which is the optimum solution, but it is unproven.

**Hint:** Full FFBC for SREFLP seems practical for the full computation skeleton of Enhanced LB more than Non-Full SREFLP, whose some ($n(n$-$1)/2$) total frequencies are zeros. Although Enhanced LB outperforms Bi-directional LB results (Shown in **Table 1**), it is still the least minimum (primal) LB and the best time complexity $O(n^2)$ for FFBC Problem regarding literature (Shown in **Table 2**).

### 3.1.3 Limitations

From (Comparative Example and Experimental Example) context, Enhanced LB is considered problematic as an independent-solution tool because it does not provide any gain for the algorithm (solution/s in permutation forms). In the next section, I proposed a technique for adapting SREFLP LB in support of being an independent solution tool.

| Problem Size n | Bi-directional LB (Sarker et al., 1998) | Enhanced LB |
|---|---|---|
| P-5 | 1138 | 1138 |
| P-6 | 1544 | 1547 |
| P-7 | 1943 | 1960 |
| P-8 | 2463 | 2572 |
| P-9 | 2936 | 2982 |
| P-10 | 3523 | 3570 |
| P-11 | 4077 | 4174 |
| P-12 | 4760 | 4854 |
| P-13 | 5565 | 5697 |
| P-14 | 6211 | 6395 |
| P-15 | 6927 | 7099 |
| P-16 | 7955 | 8119 |
| P-17 | 8766 | 8998 |
| P-18 | 9527 | 9861 |
| P-25 | 16615 | 17159 |
| P-30 | 22559 | 23488 |

**Table 1-** Performance of Enhanced LB versus Bi-directional LB

| Problem Size n | Bi-directional LB (Sarker et al., 1998) (Sarker, 2003)* | Enhanced LB | GLB (Palubeckis, 2012) | LB* (Palubeckis, 2012) |
|---|---|---|---|---|
| S-12 | 4008* | 4094 | 4122 | 4312 |
| S-13 | 5260* | 5384 | 5432 | 5730 |
| S-14 | 6548* | 6693 | 6748 | 7108 |
| S-15 | 8002* | 8199 | 8267 | 8702 |
| S-16 | 9870 | 10104 | 10176 | 10686 |
| S-17 | 11713 | 11979 | 12069 | 12759 |
| S-18 | 13949 | 14279 | 14378 | 15224 |
| S-21 | 21861 | 22450 | 22632 | 24059 |
| S-22 | 25095 | 25779 | 25994 | 27543 |
| S-23 | 28980 | 29757 | 29982 | 31823 |
| S-24 | 32893 | 33747 | 34015 | 36238 |
| S-25 | 37048 | 38064 | 38355 | 40856 |

**Table 2-** Performance of Enhanced LB versus Bi-directional LB, GLB and LB*

### 4. Gilmore-Lawler Bound

Regarding (Palubeckis, 2012), GLB is tailored for SREFLP also based on FFTC as illustrated in the following numerical example.

### 4.1 Numerical Example

From Obata benchmarks, Instance O flow matrix is

| To\From | Facility 1 | Facility 2 | Facility 3 | Facility 4 | Facility 5 |
|---|---|---|---|---|---|
| Facility 1 | 0 | 1 | 5 | 5 | 7 |
| Facility 2 | 1 | 0 | 8 | 3 | 4 |
| Facility 3 | 5 | 8 | 0 | 1 | 5 |
| Facility 4 | 5 | 3 | 1 | 0 | 7 |
| Facility 5 | 7 | 4 | 5 | 7 | 0 |

Table 3- From-To Chart for Instance O

*Step 1:* Computing the lower bound on the part of the QAP objective function corresponding to the placement of facility 1 to all positions as follows: The descending sequence of flows of facility 1 = [ 7 5 5 1], then the ascending sequence of distance *(j-i)* for Location 1 and 5 = [ 1 2 3 4], Location 2 and 4 = [ 1 1 2 3], and Location 3 = [1 1 2 2].

*Step 2:* Taking the component wise product of these sequences. Therefore, 7*1+5*2+5*3+1*4=36 is the lower bound imposed by assignment facility 1 to location 1. Other values of facility 1 locations are calculating in the same way. Also, facility 2,3,4, and 5 for every location.

*Step 3:* Constructing a linear assignment matrix for all O-5 lower bounds values where every row represents the placement of every facility to all positions P(1) to P(5).

|  | Facility 1 | Facility 2 | Facility 3 | Facility 4 | Facility 5 |
|---|---|---|---|---|---|
| P(1) | 36 | 29 | 37 | 30 | 52 |
| P(2) | 25 | 21 | 26 | 21 | 36 |
| P(3) | 24 | 20 | 25 | 20 | 32 |
| P(4) | 25 | 21 | 26 | 21 | 36 |
| P(5) | 36 | 29 | 37 | 30 | 52 |

Table 4- Linear Assignment Problem for Instance O

*Step 4:* By calculating the optimal solution of linear assignment matrix. The optimal value is 142 which is the GL bound for O-5.

### 4.2 Limitations

As observed, SREFLP is now tailored to be solved by GLB. Now, the question is: "how to guarantee optimum assignment permutations by SREFLP LB?". In particular, **"how to satisfy feasible assignment permutations by GLB?"**

## 5. Integration between Enhanced LB and Gilmore-Lawler Bound

In order to search the gap between the functionality of LB as a measuring/aid tool and as a potential solution tool. We studied the possibility of integrating the Enhanced LB and the well-known GLB. Accordingly, we developed the following rationale proof of an Optimum Scenario for SREFLP.

### 5.1 Rationale Proof for SREFLP Optimum Scenario

**Lemma 1** (OPT-Scenario)

Suppose GLB = Enhanced LB for an SREFLP instance, it can be concluded that the optimal solution (permutation matrix) follows either **Sequencing Strategy 1** = P(1), P(n), P(n-1), P(2), P(n-2), P(3), P(n-3).. etc. or **Sequencing Strategy 2** = P(1), P(n), P(2), P(n-1), P(3), P(n-2), P(4).. etc. and both if they agree in the objective function value.

**Proof**

**Sequencing Strategies 1** and **2**, corresponding to Flow Matrix term $f$ and Distance Matrix term $(j-i)$, are implicitly considered the best sequencing scenario of positioning $f$ into $(j-i)$ according to the objective function Skelton for SREFLP in **(1)**.

Regarding Enhanced LB, it is computed by taking the wise-product of both descending flow components array and ascending distance components array. While GLB entries (in the form of a linear assignment problem) are computed as in the product of Enhanced LB but wisely in separate way. It is done by taking every lower bound on the part of the QAP objective function in equation (1) for From-To Chart corresponding to the assignment of $n$ facilities per every position (the diagonal zero entries are ignored). The value of GLB is the optimal solution to the Linear Assignment Problem LAP.

*The rationale* behind the OPT-Scenario of Enhanced LB = GLB can be deeply interpreted by the assignment procedure of Optimal Permutation/s with the potential feasible minimum objective function as follows:

1) the minimum lower bound of the worst-case distance matrix part (which is related to Shipping and Receiving Facility Positions = [ 1 2 3 4 .. n-1]).
2) the minimum lower bound of the following worst-case distance matrix part (which is related to Positions P(2) and P(n-1) = [1 1 2 3 4 .. n-2]) ...etc.

In addition, from the part $(j-i)$ of From-To Chart, it can be noticed that Sequencing Strategies follow a mirror assignment for the positions corresponding to the distance part and then as a result also for the flow part shown in the wise product value in LAP.

**Hint:** Sequencing Strategies (in **Lemma 1**) aim to efficiently reduce the trials of LAP optimal solution into only two possible optimum ones.

## 5.2 Assignment of Optimum Shipping/Receiving Facility

As a result of **lemma 1**, I found that assignment of the minimum Lower Bound on the part of the QAP objective function corresponding to the first Position P(1) rationally always returns an Optimum Shipping/Receiving Facility Layout/s (shown in **Fig. 3.**). That is reviewed by (Atta & Sinha Mahapatra, 2019) from the table of SREFLP best layouts.

**Hint:** For SREFLP, this result agreed with Symmetry Breaking Constrain mentioned in (Shabani et al., 2020). It is stated that: the objective value of sequence (1 2 3 4) is equal to the one from (4 3 2 1). Therefore, as long as only the first facility of sequence/s is guaranteed to be assigned, there is no clue if it is the Shipping or Receiving one.

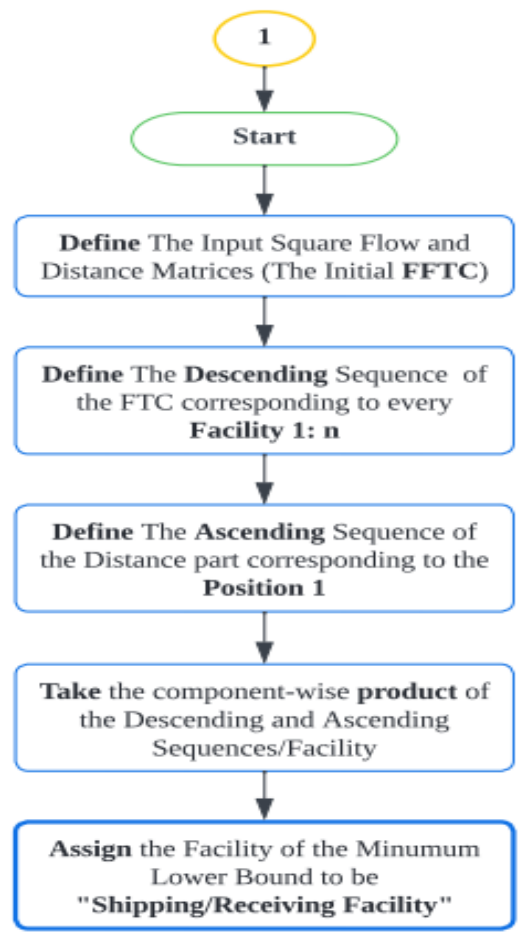

**Fig. 3.** Flow Chart of Exact Assignment of Shipping/Receiving Facility

### 5.3 Conjecture on SREFLP Approximation Scheme

We modified the assignment algorithm mentioned in **Fig. 4.** followed by Five Sequencing Strategies SSs proposed to be merged into the Exact Assignment of Shipping/Receiving Facility. SSs are assigned as follows:

**Strategy 0:** P(1), P(2), P(3), P(4), .., P(n-1), P(n).
**Strategy 1:** P(1), P(n), P(n-1), P(2), P(n-2), P(3), P(n-3)..etc.
**Strategy 2:** P(1), P(n), P(2), P(n-1), P(3), P(n-2)..etc.
**Strategy 3:** P(1), P(n), P[ 2:1:n-1].
**Strategy 4:** P(1), P(n), P[ n-1:-1:2].

**Table 5-** The Proposed Sequencing Strategies for Facilities Assignment

According to **Fig. 4.**, SREFLP family of five algorithms (also called Optimum Shipping/Receiving Facility scheme) has a running time strongly polynomial in (*n*) and is computationally performed $O(n^2 \log(n))$. Hence, SREFLP scheme is assumed to be fully input polynomial-time approximation class as defined in (van Leeuwen and van Leeuwen 2012), it is unproven.

Experimental analysis claims that the values of objective function for all benchmarks results -from optimal assignment of the Shipping/Receiving Facility- are equal to at most 4/3 the optimal solution (see **Fig. 5.**). It is available to see all Full FFBC benchmarks tested by Optimum Shipping/Receiving Facility scheme through: https://SHROUQGAMAL.github.io/OPT/. The optimal value/s OPT for all tested benchmarks are known and reported from (Palubeckis 2012).

### 5.4 Further Related Comments

In addition to the two Optimum Scenarios of Eigenvalue Related Bound - mentioned in (Meskar & Eshghi, 2020), this paper is to demonstrate and collect the whole study related to optimum SREFLP scenarios in support of P is being equal to NP. Regarding the superior LB in the literature, LB* in (Palubeckis, 2012) and improved LB* in (Hungerländer, 2014) aim implicitly to maximize GLB to be a tighter bound for various solution approaches. As a result, they served the non-polynomial exact/approximate solutions and did not serve the thrust research direction of this paper toward P=NP.

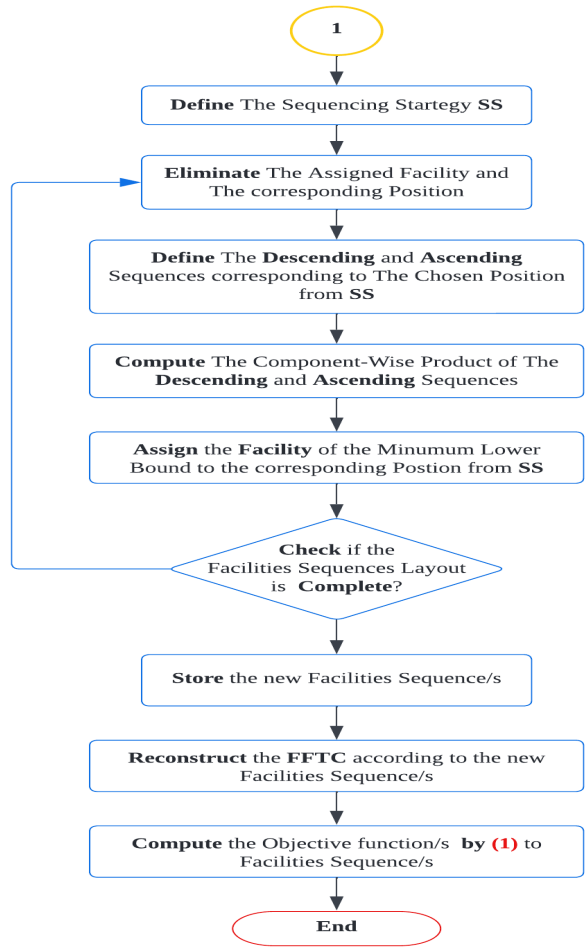

**Fig. 4.** Flow Chart of the Proposed SREFLP Approximation Scheme

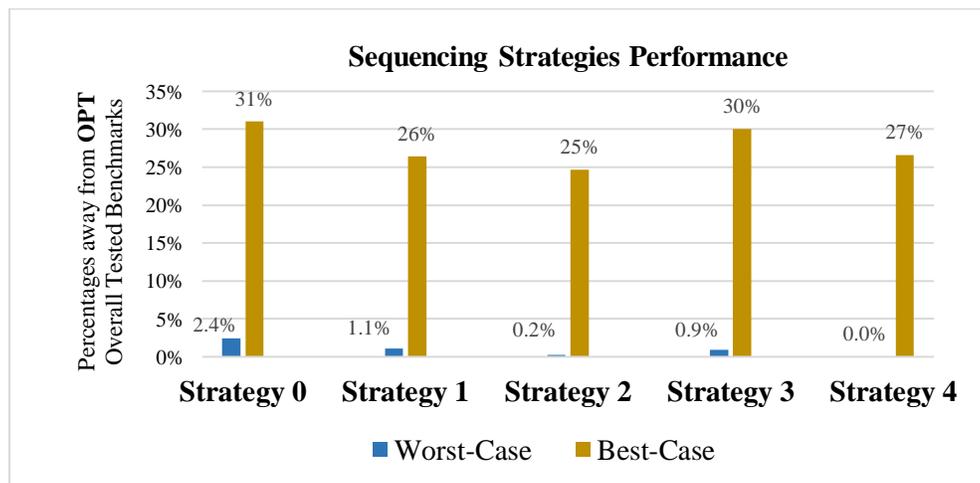

**Fig. 5.** Recap of experimental analysis for the conjecture on SREFLP Approximation Scheme

# 6. Conclusion and Future Works

In this paper we adapted the SREFLP for the first time in support of P is equal to NP. Considering the polynomial time and the exact solution, we proposed an Optimum Scenario for SREFLP and an optimum Shipping/Receiving Facility for the layout. Considering the polynomial-time and the quality of solutions, we established a conjuncture on a family of five fully input polynomial-time approximation algorithms returned objective functions equals at most 4/3 the optimal objective function of SREFLP. That was seen absolutely for the full SREFLP.

There are some interesting directions and open questions for SREFLP further research:

- Is there a proof on the conjuncture of the 4/3 fully input polynomial-time approximation scheme for full SREFLP?
- Can it apply also to the non-full SREFLP?
- Can we improve the approximation factor of SREFLP scheme?
- Is there a fixed approximability for the nature of SREFLP skeleton?